\documentclass[12pt,twoside]{article}
\usepackage{graphicx}
\usepackage{amssymb}
\usepackage{amsmath}
\begin{document}
\setcounter{page}{1}
\newtheorem{t1}{Theorem}[section]
\newtheorem{d1}{Definition}[section]
\newtheorem{c1}{Corollary}[section]
\newtheorem{l1}{Lemma}[section]
\newtheorem{r1}{Remark}[section]

\newcommand{\cA}{{\cal A}}
\newcommand{\cB}{{\cal B}}
\newcommand{\cC}{{\cal C}}
\newcommand{\cD}{{\cal D}}
\newcommand{\cE}{{\cal E}}
\newcommand{\cF}{{\cal F}}
\newcommand{\cG}{{\cal G}}
\newcommand{\cH}{{\cal H}}
\newcommand{\cI}{{\cal I}}
\newcommand{\cJ}{{\cal J}}
\newcommand{\cK}{{\cal K}}
\newcommand{\cL}{{\cal L}}
\newcommand{\cM}{{\cal M}}
\newcommand{\cN}{{\cal N}}
\newcommand{\cO}{{\cal O}}
\newcommand{\cP}{{\cal P}}
\newcommand{\cQ}{{\cal Q}}
\newcommand{\cR}{{\cal R}}
\newcommand{\cS}{{\cal S}}
\newcommand{\cT}{{\cal T}}
\newcommand{\cU}{{\cal U}}
\newcommand{\cV}{{\cal V}}
\newcommand{\cX}{{\cal X}}
\newcommand{\cW}{{\cal W}}
\newcommand{\cY}{{\cal Y}}
\newcommand{\cZ}{{\cal Z}}

\def\cl{\centerline}
\def\bd{\begin{description}}
\def\be{\begin{enumerate}}
\def\ben{\begin{equation}}
\def\benn{\begin{equation*}}
\def\een{\end{equation}}
\def\eenn{\end{equation*}}
\def\benr{\begin{eqnarray}}
\def\eenr{\end{eqnarray}}
\def\benrr{\begin{eqnarray*}}
\def\eenrr{\end{eqnarray*}}
\def\ed{\end{description}}
\def\ee{\end{enumerate}} 
\def\al{\alpha}
\def\b{\beta}
\def\bR{\bar\R}
\def\bc{\begin{center}}
\def\ec{\end{center}}
\def\d{\dot}
\def\D{\Delta}
\def\del{\delta}
\def\ep{\epsilon}
\def\g{\gamma}
\def\G{\Gamma}
\def\h{\hat}
\def\iny{\infty}
\def\La{\Longrightarrow}
\def\la{\lambda}
\def\m{\mu}
\def\n{\nu}
\def\noi{\noindent}
\def\Om{\Omega}
\def\om{\omega}
\def\p{\psi}
\def\pr{\prime}
\def\r{\ref}
\def\R{{\bf R}}
\def\ra{\rightarrow}
\def\s{\sum_{i=1}^n}
\def\t{\tau}
\def\th{\theta}
\def\Th{\Theta}

\def\vep{\varepsilon}
\def\vp{\varphi}
\def\pa{\partial}
\def\un{\underline}
\def\ov{\overline}
\def\fr{\frac}
\def\sq{\sqrt}

\def\WW{\begin{stack}{\circle \\ W}\end{stack}}
\def\ww{\begin{stack}{\circle \\ w}\end{stack}}
\def\st{\stackrel}
\def\Ra{\Rightarrow}
\def\R{{\mathbb R}}
\def\bi{\begin{itemize}}
\def\ei{\end{itemize}}
\def\i{\item}
\def\bt{\begin{tabular}}
\def\et{\end{tabular}}
\def\lf{\leftarrow}
\def\nn{\nonumber}
\def\va{\vartheta}
\def\wh{\widehat}
\def\vs{\vspace}
\def\Lam{\Lambda} 
\def\sm{\setminus}
\def\ba{\begin{array}}
\def\ea{\end{array}} 
\def\bd{\begin{description}}
\def\ed{\end{description}}
\def\lan{\langle}
\def\ran{\rangle}

\bc

\textbf{ An Experimentally accessible geometric measure for entanglement in $N$-qudit pure states}

\vspace{.2 in} 

 \textbf{ Ali Saif M. Hassan\footnote{Electronic address: alisaif@physics.unipune.ernet.in} and  Pramod S. Joag\footnote{Electronic address: pramod@physics.unipune.ernet.in}}\\
 Department of Physics, University of Pune, Pune-411007, India.
 \ec

We present a multipartite entanglement measure for $N$-qudit pure states, using the norm of the correlation tensor which occurs in the Bloch representation of the state. We compute this measure for  important class of $N$-qutrit pure states, namely general GHZ states. We prove that this measure possesses almost all the properties expected of a good entanglement measure, including monotonicity. Finally, we extend this measure to $N$-qudit mixed states via convex roof construction  and establish its various properties, including its monotonicity. \\

\section{Introduction}

\vspace{.2in}

Entanglement has proved to be a vital physical resource for various kinds of quantum information processing, including quantum state teleportation [1,2], cryptographic key distribution [3], classical communication over quantum channels [4,5,6], quantum error correction [7], quantum computational speedups [8] and distributed computation [9,10].
Further, entanglement is expected to play a crucial role in the many particle phenomena such as quantum phase transitions, transfer of information across a spin chain   [11,12] etc. Therefore, quantification of entanglement of multipartite quantum  states is fundamental to the whole field of quantum information and in general, to the physics of multicomponent quantum systems. Whereas the entanglement in pure bipartite states is well understood, classification of multipartite pure states and mixed states, according to the degree and character of their entanglement is still a matter of intense research [13,14.15]. Principal achievements are in the setting of bipartite systems. Among these, one highlights Wootter's formula for the entanglement of formation of two qubit mixed states [16], which still awaits a viable generalization to multiqubit case. Others include corresponding results for highly symmetric states [17,18,19].

Interest in multi-dimensional entangled states comes from the foundations of quantum mechanics as well as the
development of new protocols in quantum communication. For example, it has been shown that maximally entangled states of two quantum systems in a high dimensional Hilbert space, qudits, violate local realism stronger than qubits, and
entangled qudits are less affected by noise than entangled qubits [20,21]. In quantum cryptography [22], the use of entangled qutrits [22,23] or qudits [24,25] instead of qubits is more secure against eavesdropping attacks. Furthermore, the protocols for quantum teleportation or for quantum cryptography work best with maximally entangled states. These facts motivate the development of techniques to generate entangled states among quantum systems in a higher dimensional Hilbert space with good entanglement characteristics. Technical developments in this direction have been made. For example, four polarized entangled photons have been used to form two entangled
qutrits [26]. Entangled qutrits with two photons using an unbalanced 3-arm fiber optic interferometer or photonic orbital angular momentum have been demonstrated [27,28]. Time-bin entangled qudits of up to 11 dimensions from pump pulses generated by a mode-locked laser have also been reported [29]. In short, quantifying the entanglement measure of a qudit system is of physical interest. The issue of entanglement in multipartite and higher dimensional states is far more complex. Notable achievements in this area include applications of the relative entropy [30], negativity [31] Schimidt measure [32] and the global entanglement measure proposed by Meyer and Wallach [33].  

A measure of entanglement is a function on the space of states of a multipartite system, which is invariant on individual parts. Thus a complete characterization of entanglement is the characterization of all such functions. Under the most general local operations assisted by classical communication (LOCC), entanglement is expected to decrease. A measure of entanglement which decreases under LOCC is called an entanglement monotone. On bipartite pure states the sums of the $k$ smallest eigenvalues of the reduced density matrix are entanglement monotones. However, the number of independent invariants (i.e. the entanglement measures) increases exponentially as the number of particles $N$ increases and complete characterization rapidly becomes impractical. A pragmatic approach would be to seek a measure which is defined for any number of particles (scalable), which is easily calculated and which provides physically relevant information or equivalently, which passes the tests expected of a {\it good} entanglement measure [13,14].

In this paper, we present a global entanglement measure for $N$-qudit pure states which is scalable, which passes most of the tests expected of a good measure and whose value for a given system can be determined experimentally, without having a detailed {\it prior} knowledge of the state of the system. The measure is based on the Bloch representation of multipartite quantum states [34].

The paper is organized as follows. In section 2 we give the Bloch representation of a $N$-qudit quantum state and define our measure $E_{\mathcal{T}}.$ In section 3 we compute $E_{\mathcal{T}}$ for an important class of $N$-qudit states, namely, $GHZ$ states. In section 4 we prove various properties of $E_{\mathcal{T}},$ including its monotonicity, expected of a good entanglement measure. In section 5 we extend $E_{\mathcal{T}}$ to $N$-qudit mixed states via convex roof and establish its monotonicity. Finally, we summarize in section 6.

\section{Bloch representation of a $N$-partite quantum state}

Bloch representation [35,36,37,38,39] of a density operator acting on the Hilbert space of a $d$-level quantum system $\mathbb{C}^d$ is given by [40] $$ \rho = \fr{1}{d} (I_d + \sum_i s_i \lambda_i) \eqno{(1)}$$ 
 Eq.(1) is the expansion of $\rho$ in the Hilbert-Schmidt basis $\{I_d,\lambda_i; i=1,2,\dots,d^2-1\}$ where $\lambda_i$ are the traceless hermitian generators of $SU(d)$ satisfying $Tr(\lambda_i \lambda_j)=2\delta_{ij}$
  and are characterized by the structure constants of the corresponding Lie algebra, $f_{ijk}, g_{ijk}$ which are,
     respectively, completely antisymmetric and completely symmetric.
 $$\lambda_i \lambda_j =\fr{2}{d} \delta_{ij} I_d + i f_{ijk}\lambda_k +g_{ijk}\lambda_k    \eqno{(2)}$$
 
 $\textbf{s} = (s_1,s_2,\dots,s_{d^2-1})$ in Eq.(1) are the vectors in $\mathbb{R}^{d^2-1}$, constrained by the positive semidefiniteness of $\rho$,  called Bloch vectors [38]. The set of all Bloch vectors that constitute a density operator is known as the Bloch vector space $B(\mathbb{R}^{d^2-1})$. The problem of determining $B(\mathbb{R}^{d^2-1})$ where  $d\ge 3$ is still open [36,37]. However, for pure states $(\rho=\rho^2)$ the following relations hold.
 $$||\textbf{s}||_2 = \sqrt\fr{d(d-1)}{2};\;\;\; s_i s_j g_{ijk}=(d-2)s_k     \eqno{(3)}$$
  where $||.||_2$ is the Euclidean norm in $\mathbb{R}^{d^2-1}$.
  
  It is Known [41,42] that $B(\mathbb{R}^{d^2-1})$ is a subset of the ball $D_R(\mathbb{R}^{d^2-1})$ of radius $R=\sq{\fr{d(d-1)}{2}}$ , which is the minimum ball containing it, and that the ball $D_r(\mathbb{R}^{d^2-1})$ of radius $r=\sq{\fr{d}{2(d-1)}}$ is included in $B(\mathbb{R}^{d^2-1})$. That is, 
  
  $$D_r(\mathbb{R}^{d^2-1})\subseteq B(\mathbb{R}^{d^2-1}) \subseteq D_R(\mathbb{R}^{d^2-1}) \eqno{(4)}$$
  
  In order to give the Bloch representation of a density operator acting on the Hilbert   
  space $\mathbb{C}^{d} \otimes \mathbb{C}^{d} \otimes \cdots \otimes \mathbb{C}^{d}$
  of a $N$-qudit quantum system, we introduce following notation. We use $k$, \; $k_i \; (i=1,2,\cdots)$ to denote a qudit chosen from $N$ qudits, so that $k$,\; $k_i \; (i=1,2,\cdots)$ take values in the set  $\mathcal{N}=\{1,2,\cdots,N\}$. The variables $\alpha_k \;\mbox{or} \; \alpha_{k_i}$ for a given $k$ or $k_i$ span the set of generators of $SU(d)$ group (Eqs.(1) and (2)) for the $k$th or $k_i$th qudit, namely the set $\{\la_1,\la_2,\cdots,\la_{{d}^2-1}\}$ for the $k_i$th qudit. For two qudits $k_1$ and $k_2$ we define
  
   $$\lambda^{(k_1)}_{\alpha_{k_1}}=(I_{d}\otimes I_{d}\otimes \dots \otimes \lambda_{\alpha_{k_1}}\otimes I_{d}\otimes \dots \otimes I_{d})   $$
   $$\lambda^{(k_2)}_{\alpha_{k_2}}=(I_{d}\otimes I_{d}\otimes \dots \otimes \lambda_{\alpha_{k_2}}\otimes I_{d}\otimes \dots \otimes I_{d})  $$
   $$\lambda^{(k_1)}_{\alpha_{k_1}} \lambda^{(k_2)}_{\alpha_{k_2}}=(I_{d}\otimes I_{d}\otimes \dots \otimes \lambda_{\alpha_{k_1}}\otimes I_{d}\otimes \dots \otimes \lambda_{\alpha_{k_2}}\otimes I_{d}\otimes I_{d})   \eqno{(5)}$$
   
  where  $\lambda_{\alpha_{k_1}}$ and $\lambda_{\alpha_{k_2}}$ occur at the $k_1$th and $k_2$th places (corresponding to $k_1$th and $k_2$th qudits respectively) in the tensor product and are the $\alpha_{k_1}$th and  $\alpha_{k_2}$th generators of $SU(d),\; \alpha_{k_1}=1,2,\dots,d^2-1\; \mbox{and} \; \alpha_{k_2}=1,2,\dots,d^2-1$ respectively. Then we can write

$$\rho=\fr{1}{d^N} \{ I_{d}^{\otimes^N}+ \sum_{k \in \mathcal{N}}\sum_{\alpha_{k}}s_{\alpha_{k}}\lambda^{(k)}_{\alpha_{k}} +\sum_{\{k_1,k_2\}}\sum_{\alpha_{k_1}\alpha_{k_2}}t_{\alpha_{k_1}\alpha_{k_2}}\lambda^{(k_1)}_{\alpha_{k_1}} \lambda^{(k_2)}_{\alpha_{k_2}}+\cdots +$$
$$\sum_{\{k_1,k_2,\cdots,k_M\}}\sum_{\alpha_{k_1}\alpha_{k_2}\cdots \alpha_{k_M}}t_{\alpha_{k_1}\alpha_{k_2}\cdots \alpha_{k_M}}\lambda^{(k_1)}_{\alpha_{k_1}} \lambda^{(k_2)}_{\alpha_{k_2}}\cdots \lambda^{(k_M)}_{\alpha_{k_M}}+ \cdots+\sum_{\alpha_{1}\alpha_{2}\cdots \alpha_{N}}t_{\alpha_{1}\alpha_{2}\cdots \alpha_{N}}\lambda^{(1)}_{\alpha_{1}} \lambda^{(2)}_{\alpha_{2}}\cdots \lambda^{(N)}_{\alpha_{N}}\}.\eqno{(6)}$$

 where $\textbf{s}^{(k)}$ is a Bloch vector corresponding to $k$th qudit, $\textbf{s}^{(k)} =[s_{\alpha_{k}}]_{\alpha_{k}=1}^{d^2-1} $ which is a tensor of order one defined by
 $$s_{\alpha_{k}}=\fr{d}{2} Tr[\rho \lambda^{(k)}_{\alpha_{k}}]= \fr{d}{2} Tr[\rho_k \lambda_{\alpha_{k}}],\eqno{(7a)}$$ where $\rho_k$ is the reduced density matrix for the $k$th qudit. Here $\{k_1,k_2,\cdots,k_M\},\; 2 \le M \le N,$ is a subset of $\mathcal{N}$ and can be chosen in $\binom{N}{M}$  ways, contributing $\binom{N}{M}$ terms in the sum $\sum_{\{k_1,k_2,\cdots,k_M\}}$ in Eq.(6), each containing a tensor of order $M$. The total number of terms in the Bloch representation of $\rho$ is $2^N$. We denote the tensors occurring in the sum $\sum_{\{k_1,k_2,\cdots,k_M\}},\; (2 \le M \le N)$ by $\mathcal{T}^{\{k_1,k_2,\cdots,k_M\}}=[t_{\alpha_{k_1}\alpha_{k_2}\cdots \alpha_{k_M}}]$ which  are defined by 
 
 $$t_{\alpha_{k_1}\alpha_{k_2}\dots\alpha_{k_M}}=\fr{d^M}{2^M} Tr[\rho \lambda^{(k_1)}_{\alpha_{k_1}} \lambda^{(k_2)}_{\alpha_{k_2}}\cdots \lambda^{(k_M)}_{\alpha_{k_M}}]$$
 
$$ =\fr{d^M}{2^M} Tr[\rho_{k_1k_2\dots k_M} (\lambda_{\alpha_{k_1}}\otimes\lambda_{\alpha_{k_2}}\otimes\dots \otimes\lambda_{\alpha_{k_M}})]   \eqno{(7b)}$$

where $\rho_{k_1k_2\dots k_M}$ is the reduced density matrix for the subsystem $\{k_1 k_2\dots k_M\}$. Each of the  $\binom{N}{M}$ tensors of order $M$, occurring in the Bloch representation of $\rho$, contains all information about entanglement of the corresponding set of $M$ subsystems. All information on the entanglement contained in $\rho$ is coded in the tensors occurring in the Bloch representation of $\rho$. The tensor in last term in Eq. (6), we call it  $\mathcal{T}^{(N)}$, contains all the  information of genuine $N$-partite entanglement. \\

The following we give the  generators of $SU(3)$ in the $|1\ran,\;|2\ran,\;|3\ran$ basis for later use.\\
$\lambda_1=|1\ran\lan 2|+|2\ran \lan 1|$\\
$\lambda_2=-i(|1\ran\lan 2|-|2\ran \lan 1|)$\\
$\lambda_3=|1\ran\lan 1|-|2\ran \lan 2|$\\
$\lambda_4=|1\ran\lan 3|+|3\ran \lan 1|$\\
$\lambda_5=-i(|1\ran\lan 3|-|3\ran \lan 1|)$\\
$\lambda_6=|2\ran\lan 3|+|3\ran \lan 2|$\\
$\lambda_7=-i(|2\ran\lan 3|-|3\ran \lan 2|)$\\
$\lambda_8=\fr{1}{\sqrt{3}}(|1\ran\lan 1|+|2\ran \lan 2|-2 |3\ran\lan3|)$\\

The action of these generators on the basis states $\{|1\ran,|2\ran,|3\ran\}$ is given by

$\lambda_1 |1\ran=|2\ran,\; \lambda_1 |2\ran=|1\ran,\; \lambda_1 |3\ran=0$\\
$\lambda_2 |1\ran=i|2\ran,\; \lambda_2 |2\ran=-i|1\ran,\; \lambda_2 |3\ran=0$\\
$\lambda_3 |1\ran=|1\ran,\; \lambda_3 |2\ran=-|2\ran,\; \lambda_3 |3\ran=0$\\
$\lambda_4 |1\ran=|3\ran,\; \lambda_4 |2\ran=0,\; \lambda_4 |3\ran=|1\ran$\\
$\lambda_5 |1\ran=i|3\ran,\; \lambda_1 |2\ran=0,\; \lambda_1 |3\ran=-i|1\ran$\\
$\lambda_6 |1\ran=0,\; \lambda_6 |2\ran=|3\ran,\; \lambda_6 |3\ran=|2\ran$\\
$\lambda_7 |1\ran=0,\; \lambda_7 |2\ran=i|3\ran,\; \lambda_7 |3\ran=-i|2\ran$\\
$\lambda_8 |1\ran=\fr{1}{\sqrt{3}}|1\ran,\; \lambda_8 |2\ran=\fr{1}{\sqrt{3}}|2\ran,\; \lambda_8 |3\ran=-\fr{2}{\sqrt{3}}|3\ran$\\
 
 We will use these equations below (see the next section and proof of proposition 7).

We propose the following measure for a $N$-qudit pure state entanglement $$E_{\mathcal{T}}(|\psi\ran)=||\mathcal{T}^{(N)}||-(\fr{d(d-1)}{2})^{(N/2)} \eqno{(8)}$$

where $\mathcal{T}^{(N)}$ is given by Eq.(7b) for ($M=N$) in Bloch representation of $\rho=|\psi\ran \lan\psi|$. The norm of the tensor $\mathcal{T}^{(N)}$ appearing in definition (8) is the Hilbert-Schmidt (Euclidean) norm $||\mathcal{T}^{(N)}||^2=(\mathcal{T}^{(N)},\mathcal{T}^{(N)})= \sum_{\alpha_{1}\alpha_{2}\cdots \alpha_{N}}t_{\alpha_{1}\alpha_{2}\cdots \alpha_{N}}^2$. Throughout this paper, by norm, we mean the Hilbert-Schmidt (Euclidean) norm.\
\section{ Testing out $E_{\mathcal{T}}(|\psi\ran)$  for qutrit pure states}

Before proving various properties of $E_{\mathcal{T}}(|\psi\ran)$  we calculate it for 3- and 2-qutrit pure states employed to study multipartite entanglement in the literature. The most important 3-qutrit state is
$$|\psi\ran = \alpha |111\ran+\beta |222\ran+\gamma |333\ran, \; \alpha^2+\beta^2+\gamma^2=1 \eqno{(9)}$$ 
where $\{|1\ran,|2\ran,|3\ran\}$ is the computational basis for each qutrit. This state and its 2-qutrit restriction
$$|\psi\ran = \alpha |11\ran+\beta |22\ran+\gamma |33\ran, \; \alpha^2+\beta^2+\gamma^2=1 \eqno{(10)}$$ 
are used in various ways to understand qutrit entanglement. It is shown to break Clauser-Horn-Bell type of inequality for three qutrits [27]. It is used in protocols for quantum key distribution based on encoding in qutrit systems [23]. The 2-qutrit version is experimentally prepared [28,29]. Below we get a closed form expression for $E_{\mathcal{T}}$ for 3-qutrit state, Eq.(9) and we relate $E_{\mathcal{T}}$ for the 2-qutrit state, Eq.(10), with 2-qutrit concurrence [43].

For the state in equation (9), the general element of $\mathcal{T}^N$ is given by  
$$t_{i_1i_2\cdots i_N}=\fr{3^N}{2^N} \lan \psi| \lambda_{i_{1}} \lambda_{i_{2}}\cdots \lambda_{i_{N}}|\psi\ran \; i_k=1,2,\cdots,8,\; k=1,2,3.$$ From the action of $\lambda$ operators on the basis states as shown in the previous section, there are only  20 nonzero elements of $\mathcal{T}^N$. We have $t_{888}=\fr{1}{3^{3/2}}[\alpha^2+\beta^2+(-2)^3 \g^2].$ 
$ t_{111}=2\alpha \beta,$ $t_{333}=\alpha^2-\beta^2,$
$t_{444}=2\alpha\gamma,$ 
$t_{666}=2\beta\gamma.$ Other non-zero elements correspond to two $\lambda_2$s and $\lambda_1$; two $\lambda_5$s and $\lambda_4$ and two $\lambda_7$s and $\lambda_6.$  Further, $t_{833}=t_{383}=t_{338}=\fr{1}{\sqrt{3}}[\alpha^2+\beta^2]$ and $t_{838}=t_{388}=t_{883}=\fr{1}{3}[\alpha^2-\beta^2].$ Note that for $\alpha=\beta=\g =\fr{1}{\sqrt{3}}$ we have only $16$ non-zero elements. Thus the expression for $E_{\mathcal{T}}$ becomes 

$$E_{\mathcal{T}}(|\psi\ran)=||\mathcal{T}^{(N)}||-3^{3/2}=\fr{\sqrt{27}}{2}[27(\alpha^2\beta^2+\alpha^2\g^2+\beta^2\g^2)+\fr{9}{4}(\alpha^2-\beta^2)^2+\fr{27}{16}(\alpha^2+\beta^2)^2$$
$$+\fr{1}{16}(\alpha^2+\beta^2-8\g^2)^2]^{1/2}-\sqrt{27}
.\eqno{(11)}$$
The value of $E_{\mathcal{T}}$ for $\alpha=\beta=\g=\fr{1}{\sqrt{3}}$ is 3.0196.

\begin{figure}[!ht]
\begin{center}
\includegraphics[width=10cm,height=8cm]{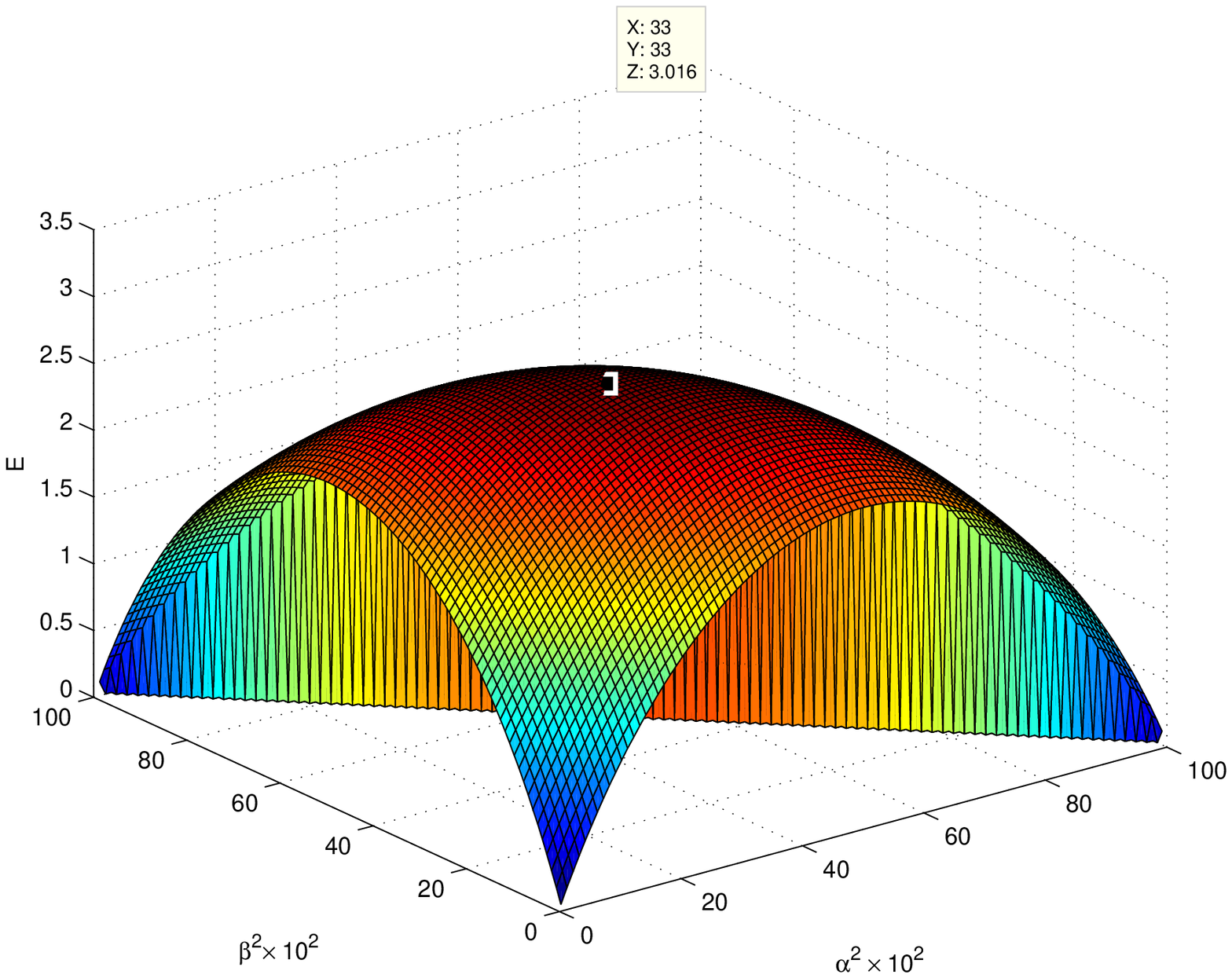}

FIG. 1. Variation of $E_{\mathcal{T}}(|GHZ\ran)$ with $\alpha^2$ and $\beta^2$. 
\end{center}
\end{figure}

The N-qutrit state corresponding to Eq.(9) for $N$-qutrits is $$|\psi\ran = \alpha |111\cdots 1\ran+\beta |222\cdots 2\ran+\gamma |333\cdots 3\ran \; \alpha^2+\beta^2+\gamma^2=1 \eqno{(12)}$$

Then any element of $\mathcal{T}^{(N)}$ namely $$t_{i_1i_2\dots i_N}=\fr{3^N}{2^N} \lan \psi| \lambda_{i_{1}} \lambda_{i_{2}}\cdots \lambda_{i_{N}}|\psi\ran, \; i_k=1,2,\cdots,8,\; k=1,\cdots,N \eqno{(13)}$$ which has $2k\;\lambda_2$s and $(N-2k)\; \lambda_1$s $k=0,1,\cdots,\lfloor \fr{N}{2} \rfloor$, where $\lfloor x \rfloor$ is the greatest integer less than or equal to $x$, is equal to $(-1)^k 2\alpha \beta$. If the element has $2k\;\lambda_5$s and $(N-2k)\; \lambda_4$s $k=0,1,\cdots,\lfloor \fr{N}{2} \rfloor$ then the value is $(-1)^k 2\alpha \g$. Similarly, if the element has $2k\; \lambda_7$s and $(N-2k)\;\lambda_6$s then the value is $(-1)^k 2 \beta\g$. Further, if the element has $2k \lambda_3$ and $(N-2k) \lambda_8$ with $k=1,2,\cdots,\lfloor \fr{N}{2} \rfloor$ then its value is $(\fr{1}{\sqrt{3}})^{N-2k} (\alpha^2+\beta^2).$ If the element has $(2k+1)\lambda_3$ and $(N-2k-1)\lambda_8$ with $k=0,1,2,\cdots ,\lfloor \fr{N}{2} \rfloor-1$ then the value is $(\fr{1}{\sqrt{3}})^{N-2k-1} (\alpha^2-\beta^2).$ Now $t_{88\cdots 8}=(\fr{1}{\sqrt{3}})^N(\alpha^2+\beta^2+(-2)^N\g^2).$ All other elements are zero. Therefore, 

$$||\mathcal{T}^N||^2=(\fr{3^N}{2^N})^2\Big{[}\sum_{k=0}^{\lfloor \fr{N}{2} \rfloor} \binom{N}{2k}(4\alpha^2\beta^2+4\alpha^2\g^2+4\beta^2\g^2)+\sum_{k=1}^{\lfloor \fr{N}{2} \rfloor}\binom{N}{2k}(\fr{1}{3})^{N-2k}(\alpha^2+\beta^2)^2$$
$$+\sum_{k=0}^{\lfloor \fr{N}{2} \rfloor-1}\binom{N}{2k+1}(\fr{1}{3})^{N-2k-1}(\alpha^2-\beta^2)^2+(\fr{1}{3})^N(\alpha^2+\beta^2+(-2)^N\g^2)^2\Big{]}$$

$$E_{\mathcal{T}}(|\psi\ran)=||\mathcal{T}^N||-3^{N/2}.\eqno{(14)}$$

We can check that $E_{\mathcal{T}}$ for $|\psi\ran$ in Eq.(14) for three qutrits is the same as $E_{\mathcal{T}}$ for $|\psi\ran$ in Eq.(11) for N qutrits with $N=3$. When $N=2$, i.e. two qutrits with $|\psi\ran$ given by Eq.(10) we get

$$E_{\mathcal{T}}(|\psi\ran)=(\fr{9}{4})[8(\alpha^2\beta^2+\alpha^2\g^2+\beta^2\g^2)+(1-\g^2)^2+(\fr{2}{3})(\alpha^2-\beta^2)^2+(\fr{1}{9})(1+3\g^2)^2]^{1/2}-3.$$

The concurrence for the state $|\psi\ran $ for two qutrits in Eq.(10)is [43] $$C(|\psi\ran)=\sqrt{4(\alpha^2\beta^2+\alpha^2\g^2+\beta^2\g^2)}.$$

Therefore, for this two qutrit state,

$$E_{\mathcal{T}}=(\fr{9}{4})[2C^2(|\psi\ran)+(1-\g^2)^2+(\fr{2}{3})(\alpha^2-\beta^2)^2+(\fr{1}{9})(1+3\g^2)^2]^{1/2}-3.\eqno{(15)}$$
We can check that whenever the concurrence $C=0,$ $E_{\mathcal{T}}=0.$

\section{ Properties of $E_{\mathcal{T}}(|\psi\ran)$}

To be a valid entanglement measure, $E_{\mathcal{T}}(|\psi\ran)$ must have the following properties [44,45].

(a) (i) {\it Positivity} : $E_{\mathcal{T}}(|\psi\ran)\ge 0$ for all $N$-qudit pure state $|\psi\ran$.
      (ii) {\it Discriminance}: $E_{\mathcal{T}}(|\psi\ran)=0$ if and only if $|\psi\ran$ is separable (product) state. 
  
(b) {\it $LU$ invariance} : $E_{\mathcal{T}}(|\psi\ran)$ is  invariant under local unitary operations.

(c) {\it Monotonicity} : local operators and classical communication ($LOCC$) do not increase the expectation value of $E_{\mathcal{T}}(|\psi\ran)$.

 We prove the above properties  for $E_{\mathcal{T}}(|\psi\ran)$. We also prove the following additional properties for $E_{\mathcal{T}}(|\psi\ran)$.
 
(d) {\it continuity} $||(|\psi\ran\lan\psi|-|\phi\ran\lan\phi|)|| \rightarrow 0 \Rightarrow \Big{|}E(|\psi\ran)-E(|\phi\ran)\Big{|}\rightarrow 0 $.
  
(e) {\it superadditivity} $E_{\mathcal{T}}(|\psi\ran \otimes |\phi\ran )\ge E_{\mathcal{T}}(|\psi\ran)+ E_{\mathcal{T}}(|\phi\ran).$\\ 
 
We need the following result which we have proved in [34].\\

\textit{ Proposition 0 }: A  pure $N$-partite quantum state is fully separable (product state) if and only if   $$\mathcal{T}^{(N)}=\mathbf{s}^{(1)}\circ \mathbf{s}^{(2)} \circ \dots \circ \mathbf{s}^{(N)}=\bigcirc_{k=1}^N s^{(k)},\eqno{(16)}$$ 
     where $ \mathbf{s}^{(k)}$ is the Bloch vector of $k$th subsystem reduced density matrix.

The symbol $\circ$ stands for  the outer product of vectors defined as follows.
  
Let $\mathbf{u}^{(1)},\mathbf{u}^{(2)},\dots,\mathbf{u}^{(M)}$ be vectors in $\mathbb{R}^{d_1^2-1},\mathbb{R}^{d_2^2-1},\cdots,\mathbb{R}^{d_M^2-1}.$ 
The outer product $\mathbf{u}^{(1)}\circ \mathbf{u}^{(2)} \circ \dots \circ \mathbf{u}^{(M)}$ is a tensor of order $M$, (M-way array), defined by

$ t_{i_1 i_2\cdots i_M}=\mathbf{u}^{(1)}_{i_1} \mathbf{u}^{(2)}_{i_2} \dots  \mathbf{u}^{(M)}_{i_M};\; 1\le i_k \le d_k^2-1,\; k=1,2,\cdots,M.$

\textit{ Proposition 1 }: Let $|\psi\ran$ be a $N$-qudit pure state. Then, $||\mathcal{T}^{(N)}_{\psi}||=(d(d-1)/2)^{(N/2)}$ if and only if $|\psi\ran$ is a separable (product) state.

\textit{Proof.} By proposition 0, $|\psi\ran$ is separable (product) if and only if $$\mathcal{T}^{(N)}=\mathbf{s}^{(1)}\circ \mathbf{s}^{(2)} \circ \dots \circ \mathbf{s}^{(N)},$$ 

As shown in [46,47 ], $$( \bigcirc_{k=1}^N s^{(k)},\bigcirc_{k=1}^N s^{(k)})=\Pi_{k=1}^N ( s^{(k)},s^{(k)}),\eqno{(17)}$$

where $(,)$ denotes the scaler product. This immediately gives, for qudits, $$||\mathcal{T}^{(N)}||^2=( \mathcal{T}^{(N)},\mathcal{T}^{(N)}) =\Pi_{k=1}^N ( s^{(k)},s^{(k)})= \Pi_k ||s^{(k)}||^2=(\fr{d(d-1)}{2})^N$$ 

Proposition 1 immediately gives

\textit{ Proposition 2 } Let $|\psi\ran$ be a $N$-qudit pure state. Then $E_{\mathcal{T}}(|\psi\ran)= 0$ if and only if $|\psi\ran$ is a product state.

\textit{Proposition 3 }: Let $|\psi\ran$ be a $N$-qudit pure state. Then $ ||\mathcal{T}^{(N)}||\ge (d(d-1)/2)^{(N/2)}.$

If $|\psi\ran$ is not a product of $N$ single qudit states (i.e. $|\psi\ran$ is not $N$-separable) then it is $N-k$ separable $k=2,3,\cdots,N-1$. Viewing the $N$-qudit system as a system comprising $N-k$ qudits, each with Hilbert space of of dimension $d$ and $k$ entangled qudits with the Hilbert space of dimension $d^k$, we can apply proposition 0 to this separable system of $N-k+1$ parts in the state $|\psi\ran.$ We get $ \mathcal{T}^{(N)}_{|\psi\ran}= \mathbf(s^{(1)})\circ\mathbf(s^{(2)})\circ \cdots\mathbf(s^{(N-k)})\circ\mathbf(s^{(N-k+1)})$

This implies, as in proposition 1, via Eq. (17) and Eq. (3) that $$||\mathcal{T}^{(N)}_{|\psi\ran}||^2=\Pi_{i-1}^{N-k+1} ||\mathbf(s^{(i)})||^2=\fr{d_k(d_k-1)}{2} (\fr{d(d-1)}{2})^{N-k} > (\fr{d(d-1)}{2})^N \; (d_k=d^k).\eqno{(18)}$$
If $k=N$ we attach an ancilla qudit in an arbitrary state $|\phi\ran$ and apply proposition 0 to $N+1$ qudit system in the state $|\psi\ran\otimes|\phi\ran$ where $|\psi\ran$ is the $N$-qudit entangled state. This result, combined with proposition 1,  completes the proof.

Proposition 3 immediately gives\\

\textit{Proposition 4} : $E_{\mathcal{T}}(|\psi\ran)\ge 0.$\\

We now prove that $E_{\mathcal{T}}(|\psi\ran)$ is nonincreasing under local operations and classical communication. Any such local action can be decomposed into four basic kinds of operations [48] (i) appending an ancillary system not entangled with the state of original system, (ii) performing a local unitary transformation, (iii) performing measurements and (iv) throwing away, i.e. tracing out, part of the system. It is clear that appending ancilla cannot change  
$\Arrowvert\mathcal{T}^{(N)}\Arrowvert.$ We prove that  $E_{\mathcal{T}}(|\psi\ran)$ does not increase under the remaining three local operations.

\textit{Proposition 5} : Let $U_i \; (i=1,2,\cdots,N)$ be a local unitary operator acting on the Hilbert space of $i$th qudit  $\mathcal{H}^{(i)}$.

Let $$\rho=(\otimes_{i=1}^N U_i)\rho'(\otimes_{i=1}^N U_i^{\dag})  \eqno{(19)}$$
for  density operators $\rho$ and $\rho'$ acting on $\mathcal{H}=\otimes_{i=1}^{N}\mathcal{H}^{(i)}$ and let $\mathcal{T}^{(N)}$ and $\mathcal{T'}^{(N)}$ denote the $N$ partite correlation tensors for $\rho$ and $\rho'$  respectively. Then, 

 $||\mathcal{T'}^{(N)}||=||\mathcal{T}^{(N)}||,$ so that $E_{\mathcal{T}}(\rho)=E_{\mathcal{T}}(\rho')$\\

 \textit{Proof.} Let $U$ denote a one qudit unitary operator, then, there exists an orthogonal matrix $[O_{\alpha \beta}]$ acting on $\mathbb{R}^{d^2-1}$ such that $U \lambda_{\alpha}U^{\dag}=\sum_{\beta} O_{\alpha \beta}\lambda_{\beta}$
 
 where $[O_{\alpha \beta}]$ is a real matrix satisfying $O O^T =I= O^T O$. 
 It is an element of the rotation group $O(d^2-1)$. Now consider $$t'_{i_1i_2\cdots i_N}=Tr(\rho'\lambda_{i_1}\otimes\lambda_{i_2}\otimes\cdots \otimes\lambda_{i_N})$$
 $$=Tr \big{(}\rho(\otimes_{i=1}^N U_i) \lambda_{i_1}\otimes\lambda_{i_2}\otimes\cdots \otimes\lambda_{i_N}(\otimes_{i=1}^N U_i^{\dag})\big{)}$$
 $$=Tr(\rho U_1\lambda_{i_1}U_1^{\dag}\otimes U_2\lambda_{i_2}U_2^{\dag}\otimes\cdots \otimes U_N\lambda_{i_N}U_N^{\dag})$$
 
 $$=\sum_{\alpha_1\cdots\alpha_N}Tr(\rho\lambda_{\alpha_1}\otimes\lambda_{\alpha_2}\otimes\cdots \otimes\lambda_{\alpha_N})O^{(1)}_{i_1\alpha_1}O^{(2)}_{i_2\alpha_2}\cdots O^{(N)}_{i_N\alpha_N}$$
 $$=\sum_{\alpha_1\cdots\alpha_N}t_{\alpha_1\cdots\alpha_N}O^{(1)}_{i_1\alpha_1}O^{(2)}_{i_2\alpha_2}\cdots O^{(N)}_{i_N\alpha_N}$$
 $$=(\mathcal{T}^{(N)}\times_1 O^{(1)}\times_2 O^{(2)}\cdots \times_N O^{(N)})_{i_1i_2\cdots i_N}$$

where $\times_k $ is the $k$-mode product of a tensor $\mathcal{T}^{(N)}\in \mathbb{R}^{(d^2-1)\times(d^2-1)\times \cdots (d^2-1)}$ by the orthogonal matrix $O^{(k)}\in \mathbb{R}^{(d^2-1)\times (d^2-1)}$ [46,47,49]. Therefore, 

$$\mathcal{T'}^{(N)}=\mathcal{T}^{(N)}\times_1 O^{(1)}\times_2 O^{(2)}\cdots \times_N O^{(N)}$$

By proposition 3.12 in [46] we get $$||\mathcal{T'}^{(N)}||= ||\mathcal{T}^{(N)}\times_1 O^{(1)}\times_2 O^{(2)}\cdots \times_N O^{(N)}||=||\mathcal{T}^{(N)} ||$$ \hfill $\blacksquare$\\

\textit{Proposition 6} : If a multipartite pure state $|\psi\ran$ is subjected to a local measurement on the $k$th qudit giving outcomes $i_k$ with probabilities $p_{i_k}$ and leaving residual $N$-qudit pure state $|\phi_{i_k}\ran$ then the expected entanglement $\sum_{i_k} p_{i_k} E_{\mathcal{T}}(|\phi_{i_k}\ran)$ of residual state is not greater then $ E_{\mathcal{T}}(|\psi\ran)$.

$$\sum_{i_k} p_{i_k} E_{\mathcal{T}}(|\phi_{i_k}\ran) \le E_{\mathcal{T}}(|\psi\ran). \eqno{(20)}$$

\textit{Proof.} Local measurements can be expressed as the tensor product matrix $\bar{D}=\bar{D}^{(1)}\otimes\bar{D}^{(2)}\otimes\cdots \otimes \bar{D}^{(N)}$  on the expanded coherence vector $\mathcal{T}$ [50]. The expanded coherence vector $\mathcal{T}$ is the extended correlation tensor $\mathcal{T}$ (defined below) viewed as a vector in the real space of appropriate dimension. The extended correlation tensor $\mathcal{T}$ is defined by the equation
$$\rho= \fr{1}{2^N}\sum_{i_1i_2\cdots i_N=0}^{d^2-1} \mathcal{T}_{i_1i_2\cdots i_N} \lambda_{i_1}\otimes \lambda_{i_2}\otimes\cdots\otimes \lambda_{i_N}  \eqno{(21)}$$
where $\lambda_{i_k} \in \{I,\lambda_1,\lambda_2,\cdots,\lambda_{d^2-1}\}$ are the $i_k$th local $SU(d)$ generators acting on the $k$th qudit $(\lambda_0 =I)$ and the real coefficients $\mathcal{T}_{i_1i_2\cdots i_N} $ are the components of the extended correlation tensor $\mathcal{T}.$  Eq. (6) and Eq.(21) are equivalent with $\mathcal{T}_{000\cdots 0}=1 $ , $\mathcal{T}_{i_100\cdots 0}= s^{(1)}_{i_1},\; \cdots $, $\mathcal{T}_{i_1i_2\cdots i_M 00\cdots 0} = \mathcal{T}^{\{1,2,\cdots M\}}_{i_1i_2\cdots i_M}, \cdots $
and  $\mathcal{T}_{i_1i_2\cdots i_N} =\mathcal{T}^{(N)}_{i_1i_2\cdots i_N} ,\; i_1, i_2,\cdots ,i_N \ne 0 .$ $\bar{D}^{(k)};\; k=1,2,\cdots N$ are $d^2\times d^2$ matrices. Without losing generality, we can assume the local measurements to be POVMs, in which case $\bar{D}^{(k)}= diag(1,D^{(k)})$ and the $(d^2-1)\times(d^2-1)$ matrix $D^{(k)}$ is contractive $D^{(k)T} D^{(k)} \le I$ [50]. The local POVMs acting on a $N$-qudit state $\rho$ corresponds to the map 
$\rho\longmapsto\mathcal{M}(\rho)$ given by
$$\mathcal{M}(\rho)=\sum_{i_1i_2\cdots i_N}L^{(1)}_{i_1}\otimes L^{(2)}_{i_2}\otimes \cdots \otimes L^{(N)}_{i_N}\rho L^{(1)\dag}_{i_1}\otimes L^{(2)\dag}_{i_2}\otimes \cdots \otimes L^{(N)\dag}_{i_N}$$
where $L^{(k)}_{i_k}$ are the linear, positive, trace preserving operators satisfying $\sum_{i_k}L^{(k)\dag}_{i_k}L^{(k)}_{i_k}=I$ and $[L^{(k)}_{i_k},L^{(k)\dag}_{i_k}]=0.$ The resulting correlation tensor of $\mathcal{M}(\rho)$ can be written as 

$$\mathcal{T'}^{(N)}=\mathcal{T}^{(N)}\times_1 D^{(1)}\times_2 D^{(2)}\cdots \times_N D^{(N)}$$
where $D^{(k)}$ is $(d^2-1)\times (d^2-1)$ matrix and $D^{(k)T} D^{(k)} \le I.$

Action of POVM on $k$th qudit corresponds to the map $\mathcal{M}_{k}(|\psi\ran\lan\psi|)=\sum_{i_k} M_{i_k}\rho M_{i_k}^{\dag}$ where $M_{i_k}=I \otimes \cdots L^{(k)}_{i_k}\otimes \cdots I$ , $\sum_{i_k}L^{(k)\dag}_{i_k}L^{(k)}_{i_k}=I$ and $[L^{(k)}_{i_k},L^{(k)\dag}_{i_k}]=0$
with the resulting mixed state $\sum_{i_k}p_{i_k}|\phi_{i_k}\ran\lan\phi_{i_k}|,$ where $|\phi_{i_k}\ran$ is the $N$-qudit pure state which results after the the outcome $i_k$ with probability $p_{i_k}.$

The average entanglement of this state is $$\sum_{i_k} p_{i_k} E_{\mathcal{T}}(|\phi_{i_k}\ran\lan\phi_{i_k}|)=\sum_{i_k}p_{i_k} ||\mathcal{T}^{(N)}_{|\phi_{i_k}\ran}||-(\fr{d(d-1)}{2})^{(N/2)}$$
$$=\sum_{i_k} p_{i_k}||\mathcal{T}^{(N)}_{|\psi\ran}\times_k D^{(k)}||-(\fr{d(d-1)}{2})^{(N/2)}$$
$$\sum_{i_k}p_{i_k}||D^{(k)}T_{(k)}(|\psi\ran)||-(\fr{d(d-1)}{2})^{(N/2)}$$
where, by proposition 3.7 in [46], $ D^{(k)}T_{(k)}(|\psi\ran)$ is the $k$th matrix unfolding [34] of $\mathcal{T}^{(N)}_{|\psi\ran}\times_k D^{(k)}.$
Therefore, from the definition of the Euclidean norm of a matrix, $||A||=\sq{Tr(A A^{\dag})}$  [51] we get
$$\sum_{i_k} p_{i_k}  E_{\mathcal{T}}(|\phi_{i_k}\ran\lan\phi_{i_k}|)=\sum_{i_k} p_{i_k}\big{[}Tr\big{(}D^{(k)}T_{(k)}(|\psi\ran)T_{(k)}^{\dag}(|\psi\ran)D^{(k)T}\big{)}\big{]}^{\fr{1}{2}}-(\fr{d(d-1)}{2})^{(N/2)}$$

$$=\sum_{i_k} p_{i_k} \big{[}Tr \big{(} D^{(k)T}D^{(k)}T_{(k)}(|\psi\ran)T_{(k)}^{\dag}(|\psi\ran)\big{)}\big{]}^{\fr{1}{2}}-(\fr{d(d-1)}{2})^{(N/2)}$$

 $$\le \sum_{i_k}p_{i_k}\sq{Tr\big{(}T_{(k)}(|\psi\ran)T_{(k)}^{\dag}(|\psi\ran)\big{)}}-(\fr{d(d-1)}{2})^{(N/2)}$$
 $$=||\mathcal{T}^{(N)}_{|\psi\ran}||-(\fr{d(d-1)}{2})^{(N/2)} = E_{\mathcal{T}}(|\psi\ran)$$ 
 because $D^{(k)T} D^{(k)} \le I$, and $\sum_{i_k}p_{i_k} = 1.$ We have also used the fact that Euclidean norm of a tensor equals that of any of its matrix unfoldings. \hfill $\blacksquare$\\
 
\textit{Proposition 7 :} Let $|\psi\ran$ be an $N$-qudit pure state. Let $\rho$ denote the reduced density matrix after tracing out one qudit from the state $|\psi\ran$. Then $$|| \mathcal{T}^{(N-1)}_{\rho}|| \le ||\mathcal{T}^{(N)}_{|\psi\ran}||$$
with equality only when $|\psi\ran=|\phi\ran\otimes|\chi\ran$ where $|\chi\ran$ is the state of the qudit which is traced out.\\

\textit{Proof.} We prove this for a case of qutrit states whose generalization to all qudit states is straightforward.
Consider  a N-qutrit state $|\psi\ran=a|b_1\cdots b_N\ran+b|b'_1\cdots b'_N\ran+c|b''_1\cdots b''_N\ran;\; |a|^2+|b|^2+|c|^2=1.$

Here $|b_i\ran$, $|b'_i\ran$ and $|b''_i\ran$ are the eigenstates of $\lambda_8^{(i)}$ operating on the $i$th qutrit. Now consider set $S$ of  $N$-fold tensor products of qutrit operators , namely  $S=\{\lambda_{\alpha_1}\otimes\lambda_{\alpha_2}\otimes \cdots \otimes \lambda_{\alpha_N}\},\; \alpha_1\cdots \alpha_N =1,\cdots,8.$

Choosing $\alpha_1,\cdots, \alpha_N =8$ we get $$\lambda_{8}\otimes\lambda_{8}\otimes \cdots\otimes \lambda_{8}|b_1\cdots b_N\ran=\fr{(-2)^s}{3^{N/2}} |b_1\cdots b_N\ran,$$ 
$$\lambda_{8}\otimes\lambda_{8}\otimes \cdots\otimes \lambda_{8}|b'_1\cdots b'_N\ran=\fr{(-2)^{s'}}{3^{N/2}} |b'_1\cdots b'_N\ran,$$
$$\lambda_{8}\otimes\lambda_{8}\otimes \cdots\otimes \lambda_{8}|b''_1\cdots b''_N\ran=\fr{(-2)^{s''}}{3^{N/2}} |b''_1\cdots b''_N\ran,$$
where $s,\;s',\;s''$ are the number of occurrences of state $|3\ran$ in $|b_1\cdots b_N\ran$, $|b'_1\cdots b'_N\ran$ and $|b''_1\cdots b''_N\ran$ respectively. We can replace $k$ of the $(N-s)$, $(N-s')$, $(N-s'')\lambda_8$ operators acting on the states $|1\ran$ and $|2\ran$ occurring in $|b_1\cdots b_N\ran$, $|b'_1\cdots b'_N\ran$, $|b''_1\cdots b''_N\ran$, respectively by $\lambda_3$ to get $\binom{N-s}{k}$, $\binom{N-s'}{k}$, $\binom{N-s''}{k}$ new operators which we call  $Y_k \;(k=1,2,\cdots,N-s)$, $Y'_k  \;(k=1,2,\cdots,N-s')$ and $Y''_k \;(k=1,2,\cdots,N-s'')$ respectively.

We can choose an operator from $S$, denoted $B$, such that $B|b_1\cdots b_N\ran = \pm \eta |b'_1\cdots b'_N\ran,$ where $\eta$ is determined by the eigenvalues of $\lambda_8.$
If $B$ contains $q \le N$ occurrences of operators $\lambda_1,\;\lambda_4,\; \lambda_6,$ then we can replace $k \le q$ of them by $\lambda_2,\; \lambda_5,\;\lambda_7$ operators respectively. We denote the tensor product operator obtained after $k$ such replacements by $B_k \; (B_0=B)$. We have, $B_k|b_1\cdots b_N\ran = \pm (i)^k \eta|b'_1\cdots b'_N\ran.$
We also choose an operator $E$ from $S$ such that $E|b_1\cdots b_N\ran = \pm \chi |b''_1\cdots b''_N\ran,$ 
also the operators $E_k\;(E_0=E)\;(k <q')$ such that $E_k|b_1\cdots b_N\ran = \pm (i)^k \chi|b''_1\cdots b''_N\ran.$
Further, we choose an operator $D$ in $S$ such that $D|b'_1\cdots b'_N\ran = \pm \delta |b''_1\cdots b''_N\ran,$ 
 and also $D_k\;(D_0=D)\;(k <q'')$ such that $D_k|b'_1\cdots b'_N\ran = \pm (i)^k \delta|b''_1\cdots b''_N\ran.$
We can Now calculate various tensor elements,  $$\lan b_1\cdots b_N\arrowvert\lambda_8\otimes\cdots\otimes\lambda_8\arrowvert b_1\cdots b_N\ran = \fr{(-2)^s}{3^{N/2}} $$
$$ \lan b'_1\cdots b'_N\arrowvert\lambda_8\otimes\cdots\otimes\lambda_8\arrowvert b'_1\cdots b'_N\ran=\fr{(-2)^{s'}}{3^{N/2}}$$
$$\lan b''_1\cdots b''_N\arrowvert\lambda_8\otimes\cdots\otimes\lambda_8\arrowvert b''_1\cdots b''_N\ran=\fr{(-2)^{s''}}{3^{N/2}}.$$

Replacing  $(N-s),\;(N-s'),\;(N-s'')\lambda_8$ by operators by $\lambda_3$ operators we get 

$$\lan b_1\cdots b_N\arrowvert\lambda_8\otimes\cdots\otimes\lambda_8\otimes\lambda_3\cdots\otimes\lambda_3\arrowvert b_1\cdots b_N\ran = \fr{(-2)^s}{3^{s/2}}$$
$$ \lan b'_1\cdots b'_N\arrowvert\lambda_8\otimes\cdots\otimes\lambda_8\otimes\lambda_3\cdots\otimes\lambda_3\arrowvert b'_1\cdots b'_N\ran=\fr{(-2)^{s'}}{3^{s'/2}}$$
$$\lan b''_1\cdots b''_N\arrowvert\lambda_8\otimes\cdots\otimes\lambda_8\otimes\lambda_3\cdots\otimes\lambda_3\arrowvert b''_1\cdots b''_N\ran=\fr{(-2)^{s''}}{3^{s''/2}}.$$

The symbols $q'$ and $q''$ have the same definition as that for $q$. Symbols $s'$ and $s''$ denote the number of times the state $|3\ran$ has occurred in $\arrowvert b'_1\cdots b'_N\ran$ and $\arrowvert b''_1\cdots b''_N\ran$ respectively.
The contributions of operators $B,\;B_k,\;E,\;E_k,\;D,\;D_k$ are,

$$\lan b_1\cdots b_N\arrowvert B \arrowvert b'_1\cdots b'_N\ran = \pm \eta = \lan b'_1\cdots b'_N\arrowvert B \arrowvert b_1\cdots b_N\ran$$
$$\lan b'_1\cdots b'_N\arrowvert B_k \arrowvert b_1\cdots b_N\ran = \pm (i)^{k}\eta$$
$$\lan b_1\cdots b_N\arrowvert B_k \arrowvert b'_1\cdots b'_N\ran = \pm (-i)^{k}\eta$$

$$\lan b_1\cdots b_N\arrowvert E \arrowvert b''_1\cdots b''_N\ran = \pm \chi = \lan b''_1\cdots b''_N\arrowvert E \arrowvert b_1\cdots b_N\ran$$
$$\lan b''_1\cdots b''_N\arrowvert E_k \arrowvert b_1\cdots b_N\ran = \pm (i)^{k}\chi$$
$$\lan b_1\cdots b_N\arrowvert E_k \arrowvert b''_1\cdots b''_N\ran = \pm (-i)^{k}\chi$$
also
$$\lan b'_1\cdots b'_N\arrowvert D \arrowvert b''_1\cdots b''_N\ran = \pm \delta = \lan b''_1\cdots b''_N\arrowvert B \arrowvert b'_1\cdots b'_N\ran$$
$$\lan b''_1\cdots b''_N\arrowvert D_k \arrowvert b'_1\cdots b'_N\ran = \pm (i)^{k}\delta$$
$$\lan b'_1\cdots b'_N\arrowvert D_k \arrowvert b''_1\cdots b''_N\ran = \pm (-i)^{k}\delta$$

Finally, we get , for the tensor element,$$ t_{\alpha_1\cdots\alpha_N} = \fr{3^N}{2^N}\lan \psi\arrowvert\lambda_{\alpha_1}\otimes\cdots\otimes\lambda_{\alpha_N}\arrowvert\psi\ran$$
$$ =\fr{3^N}{2^N}[ \arrowvert a\arrowvert^2 \lan b_1\cdots b_N\arrowvert\lambda_{\alpha_1}\otimes\cdots\otimes\lambda_{\alpha_N}\arrowvert b_1\cdots b_N\ran+|b|^2\lan b'_1 \cdots b'_N|\lambda_{\alpha_1}\otimes\cdots\otimes\lambda_{\alpha_N}|b'_1 \cdots b'_N\ran$$
$$+|c|^2\lan b''_1 \cdots b''_N|\lambda_{\alpha_1}\otimes\cdots\otimes\lambda_{\alpha_N}|b''_1 \cdots b''_N\ran$$
$$ +a^* b \lan b_1\cdots b_N\arrowvert\lambda_{\alpha_1}\otimes\cdots\otimes\lambda_{\alpha_N}|b'_1 \cdots b'_N\ran+a b^* \lan b'_1 \cdots b'_N|\lambda_{\alpha_1}\otimes\cdots\otimes\lambda_{\alpha_N}\arrowvert b_1\cdots b_N\ran$$
$$ +a^*c \lan b_1\cdots b_N\arrowvert\lambda_{\alpha_1}\otimes\cdots\otimes\lambda_{\alpha_N}|b''_1 \cdots b''_N\ran+a c^* \lan b''_1 \cdots b''_N|\lambda_{\alpha_1}\otimes\cdots\otimes\lambda_{\alpha_N}\arrowvert b_1\cdots b_N\ran$$
$$ +b^*c \lan b'_1\cdots b'_N\arrowvert\lambda_{\alpha_1}\otimes\cdots\otimes\lambda_{\alpha_N}|b''_1 \cdots b''_N\ran+b c^* \lan b''_1 \cdots b''_N|\lambda_{\alpha_1}\otimes\cdots\otimes\lambda_{\alpha_N}\arrowvert b'_1\cdots b'_N\ran]$$

The nonzero elements of $t_{\alpha_1 \cdots \alpha_N} $ are $t_{88\cdots 8}=\fr{1}{\sqrt{3^N}} ((-2)^s|a|^2 +(-2)^{s'} |b|^2+(-2)^{s''}|c|^2)$, $t_{88\cdots83\cdots3}=\fr{(-2)^s}{\sqrt{3^{N-k}}}|a|^2 +\fr{(-2)^{s'}}{\sqrt{3^{N-k}}}|b|^2+\fr{(-2)^{s''}}{\sqrt{3^{N-k}}}|c|^2)$,  where $k$, is  the number of $\lambda_3$s  in the element $\lambda_{\alpha_1}\otimes\cdots\otimes\lambda_{\alpha_N}\; \alpha_i=3,8$.  The elements of $\mathcal{T}^{(N)}$ corresponding to $B$ type of operators are

$ t_B=\pm \eta a b^* \pm \eta a^* b= \pm 2 \eta |a| |b| cos (\phi_a-\phi_b)$, 

 \begin{displaymath}
t_{B_k}=\pm (i)^k \eta a b^* \pm (-i)^k \eta a^* b 
 =\left\{ \begin{array}{ll}
 \pm 2 \eta |a|\; |b| cos (\phi_a-\phi_b) & \textrm{ if $k$ is even} \\

\pm 2 \eta |a|\; |b| sin (\phi_a-\phi_b) & \textrm{ if $k$ is odd} \\
\end{array} \right.
\end{displaymath}

We get $\sum _{k=0}^q \binom{q}{2k}$ elements with $cos (\phi_a-\phi_b)$ and $\sum _{k=0}^q \binom{q}{2k+1}$ elements with  $sin(\phi_a-\phi_b)$. If $q$ is odd (for the given state$|\psi\ran$) the number of cosines and the number of sines are equal. When $q$ is even the number of cosines exceeds by 1. Similarly

$ t_E=\pm \chi a b^* \pm \chi a^* b= \pm 2 \chi |a| |b| cos (\phi_a-\phi_c)$, 

 \begin{displaymath}
t_{E_k}=\pm (i)^k \chi a b^* \pm (-i)^k \chi a^* b 
 =\left\{ \begin{array}{ll}
 \pm 2 \chi |a|\; |b| cos (\phi_a-\phi_c) & \textrm{ if $k$ is even} \\

\pm 2 \chi |a|\; |b| sin (\phi_a-\phi_c) & \textrm{ if $k$ is odd} \\
\end{array} \right.
\end{displaymath}

We get $\sum _{k=0}^{q'} \binom{q'}{2k}$ elements with $cos (\phi_a-\phi_c)$ and $\sum _{k=0}^{q'} \binom{q'}{2k+1}$ elements with  $sin(\phi_a-\phi_c)$. If $q'$ is odd (for the given state$|\psi\ran$) the number of cosines and the number of sines are equal. When $q'$ is even the number of cosines exceeds by 1. 

$ t_D=\pm \delta a b^* \pm \delta a^* b= \pm 2 \delta |a| |b| cos (\phi_b-\phi_c)$, 

 \begin{displaymath}
t_{D_k}=\pm (i)^k \delta a b^* \pm (-i)^k \delta a^* b 
 =\left\{ \begin{array}{ll}
 \pm 2 \delta |a|\; |b| cos (\phi_b-\phi_c) & \textrm{ if $k$ is even} \\

\pm 2 \delta |a|\; |b| sin (\phi_b-\phi_c) & \textrm{ if $k$ is odd} \\
\end{array} \right.
\end{displaymath}

We get $\sum _{k=0}^{q''} \binom{q''}{2k}$ elements with $cos (\phi_b-\phi_c)$ and $\sum _{k=0}^{q''} \binom{q''}{2k+1}$ elements with  $sin(\phi_b-\phi_c)$. If $q''$ is odd (for the given state$|\psi\ran$) the number of cosines and the number of sines are equal. When $q''$ is even the number of cosines exceeds by 1.

Finally  we get, 

$|| \mathcal{T}^{(N)}_{|\psi\ran}||^2 = (\fr{3^N}{2^N})^2\times\fr{1}{3^N}\Big{[}\Big{(}(-2)^s|a|^2 +(-2)^{s'} |b|^2+(-2)^{s''}|c|^2\Big{)}^2$
$$+\sum^{N-s}_{k=1}\Big{(}\fr{(-2)^s}{3^{N-k}}\Big{)}^2 |a|^2 \binom{N-s}{k}+\sum^{N-s'}_{k=1}\Big{(}\fr{(-2)^{s'}}{3^{N-k}}\Big{)}^2|b|^2\binom{N-s'}{k}+\sum^{N-s''}_{k=1}\Big{(}\fr{(-2)^{s''}}{3^{N-k}}\Big{)}^2|c|^2\binom{N-s''}{k}$$
$$+ 4 \eta^2 |a|^2 |b|^2 cos^2 (\phi_a-\phi_b) \sum_{k=0}^q \binom{q}{2k}+4 \eta^2 |a|^2 |b|^2 sin^2 (\phi_a-\phi_b) \sum_{k=0}^q \binom{q}{2k+1}$$
$$+ 4 \chi^2 |a|^2 |c|^2 cos^2 (\phi_a-\phi_c) \sum_{k=0}^{q'} \binom{q'}{2k}+4 \chi^2 |a|^2 |c|^2 sin^2 (\phi_a-\phi_c) \sum_{k=0}^{q'} \binom{q'}{2k+1}$$
$$+ 4 \delta^2 |b|^2 |c|^2 cos^2 (\phi_b-\phi_c) \sum_{k=0}^{q''} \binom{q''}{2k}+4 \delta^2 |b|^2 |c|^2 sin^2 (\phi_b-\phi_c) \sum_{k=0}^{q''} \binom{q''}{2k+1}\Big{]}$$

Note that, using  $|a|^2+|b|^2+|c|^2 =1$, it is easy to see that $||\mathcal{T}^{(N)}_{|\psi\ran}|| \ge 3^{N/2}$, showing that $E_{\mathcal{T}} \ge 0.$

 Next we consider $$|\psi\ran \lan \psi|= |a|^2 \arrowvert b_1\cdots b_N  \ran  \lan b_1\cdots b_N \arrowvert + |b|^2 |b'_1 \cdots b'_N \ran \lan b'_1 \cdots b'_N |+ |c|^2 |b''_1 \cdots b''_N \ran \lan b''_1 \cdots b''_N |$$
 $$ + a b^* \arrowvert b_1\cdots b_N  \ran \lan b'_1 \cdots b'_N |+ a^* b |b'_1 \cdots b'_N \ran  \lan b_1\cdots b_N \arrowvert $$ 
$$ + a c^* \arrowvert b_1\cdots b_N  \ran \lan b''_1 \cdots b''_N | + a^* c |b''_1 \cdots b''_N \ran  \lan b_1\cdots b_N \arrowvert $$
 $$ + b c^* \arrowvert b'_1\cdots b'_N  \ran \lan b''_1 \cdots b''_N | + b^* c |b''_1 \cdots b''_N \ran  \lan b'_1\cdots b'_N \arrowvert $$
 
 and trace out the $N$th qudit to get the $N-1$ qudit reduced density matrix  
 $$\rho = |a|^2 \arrowvert b_1\cdots b_{N-1}  \ran  \lan b_1\cdots b_{N-1} \arrowvert + |b|^2 |b'_1 \cdots b'_{N-1} \ran \lan b'_1 \cdots b'_{N-1} | + |c|^2 |b''_1 \cdots b''_{N-1} \ran \lan b''_1 \cdots b''_{N-1} | $$
 $$+  a b^* \arrowvert b_1\cdots b_{N-1}  \ran \lan b'_1 \cdots b'_{N-1} | \lan b_N|b'_N\ran + a^* b |b'_1 \cdots b'_{N-1} \ran  \lan b_1\cdots b_{N-1} \arrowvert \lan b'_N|b_N\ran $$ 
$$+  a c^* \arrowvert b_1\cdots b_{N-1}  \ran \lan b''_1 \cdots b''_{N-1} | \lan b_N|b''_N\ran + a^* c |b''_1 \cdots b''_{N-1} \ran  \lan b_1\cdots b_{N-1} \arrowvert \lan b''_N|b_N\ran $$ 
$$+  b c^* \arrowvert b'_1\cdots b'_{N-1}  \ran \lan b''_1 \cdots b''_{N-1} | \lan b'_N|b''_N\ran + b^* c |b''_1 \cdots b''_{N-1} \ran  \lan b'_1\cdots b'_{N-1} \arrowvert \lan b''_N|b'_N\ran $$ 
  
 Now $$t_{\alpha_1\cdots \alpha_{N-1}}= Tr(\rho \lambda_{\alpha_1}\otimes \lambda_{\alpha_2}\otimes \cdots \lambda_{\alpha_{N-1}})
 = \arrowvert a\arrowvert^2 \lan b_1\cdots b_{N-1}\arrowvert\lambda_{\alpha_1}\otimes\cdots\otimes\lambda_{\alpha_{N-1}}\arrowvert b_1\cdots b_{N-1}\ran$$ 
$$ +|b|^2\lan b'_1 \cdots b'_{N-1}|\lambda_{\alpha_1}\otimes\cdots\otimes\lambda_{\alpha_{N-1}}|b'_1 \cdots b'_{N-1}\ran   +|c|^2\lan b''_1 \cdots b''_{N-1}|\lambda_{\alpha_1}\otimes\cdots\otimes\lambda_{\alpha_{N-1}}|b''_1 \cdots b''_{N-1}\ran $$
$$+a^* b \lan b_1\cdots b_{N-1}\arrowvert\lambda_{\alpha_1}\otimes\cdots\otimes\lambda_{\alpha_{N-1}}|b'_1 \cdots b'_{N-1}\ran  \lan b_N|b'_N\ran $$
$$+a b^* \lan b'_1 \cdots b'_{N-1}|\lambda_{\alpha_1}\otimes\cdots\otimes\lambda_{\alpha_{N-1}}\arrowvert b_1\cdots b_{N-1}\ran  \lan b'_N|b_N\ran$$ 
 $$+a^* c \lan b_1\cdots b_{N-1}\arrowvert\lambda_{\alpha_1}\otimes\cdots\otimes\lambda_{\alpha_{N-1}}|b''_1 \cdots b''_{N-1}\ran  \lan b_N|b''_N\ran $$
 $$+a c^* \lan b''_1 \cdots b''_{N-1}|\lambda_{\alpha_1}\otimes\cdots\otimes\lambda_{\alpha_{N-1}}\arrowvert b_1\cdots b_{N-1}\ran  \lan b''_N|b_N\ran$$
 $$+b^* c \lan b'_1\cdots b'_{N-1}\arrowvert\lambda_{\alpha_1}\otimes\cdots\otimes\lambda_{\alpha_{N-1}}|b''_1 \cdots b''_{N-1}\ran  \lan b'_N|b''_N\ran $$
 $$+b c^* \lan b''_1 \cdots b''_{N-1}|\lambda_{\alpha_1}\otimes\cdots\otimes\lambda_{\alpha_{N-1}}\arrowvert b'_1\cdots b'_{N-1}\ran  \lan b''_N|b'_N\ran$$
 
  We have for $N-1$ tensor product operators  $\lambda_{8}\otimes\lambda_{8}\otimes \cdots\otimes \lambda_{8}|b_1\cdots b_{N-1}\ran=\fr{(-2)^{\ell}}{3^{(N-1)/2}} |b_1\cdots b_{N-1}\ran.$

 We construct the operators $X,\;X_k,\;X',\;X'_k,\;X'',\;X''_k,\;G,\;G_k,\;Q,\;Q_k,\;Z$ and $Z_k$ corresponding to $Y,\;Y_k,\;Y',\;Y'_k,\;Y'',\;Y''_k,\;B,\;B_k,\;E,\;E_k,\; D$ and $D_k$  respectively acting on $N-1$ qudits. We then get $G|b_1\cdots b_N\ran = \pm \eta |b'_1\cdots b'_{N-1}\ran,$ 
   $ G_k |b_1 \cdots b_{N-1}\ran= \pm (i)^k \eta' |b'_1 \cdots b'_{N-1}\ran $ 
   
  $Q|b_1\cdots b_N\ran = \pm \chi' |b''_1\cdots b''_{N-1}\ran,$ 
   $ Q_k |b_1 \cdots b_{N-1}\ran= \pm (i)^k \chi' |b''_1 \cdots b''_{N-1}\ran $ 
   
    $Z|b'_1\cdots b'_N\ran = \pm \delta' |b''_1\cdots b''_{N-1}\ran,$ 
   $ Z_k |b'_1 \cdots b'_{N-1}\ran= \pm (i)^k \delta' |b''_1 \cdots b''_{N-1}\ran $ 
  
 Now, the nonzero elements of $\mathcal{T}^{(N-1)}_{\rho}$ are  
 $t_{88\cdots 8}=[(-2)^{\ell}|a|^2+(-2)^{\ell'}|b|^2+(-2)^{\ell''}|c|^2]$, where $\ell,\;\ell',\;\ell''$ are defined as $s,\;s',\;s''$ for $|\psi\ran$.

 $ t_G=\pm \eta' a b^* \lan b_N|b'_N\ran \pm \eta' a^* b \lan b'_N | b_N\ran= \pm 2 \eta' |a| |b| |\lan b'_N | b_N\ran| cos (\phi_a-\phi_b-\alpha_G)$, \\
 
 $t_{G_k}=\pm (i)^k \eta' a b^*  \lan b_N|b'_N\ran \pm (-i)^k \eta' a^* b \lan b'_N | b_N\ran$ \\ 
  
 \begin{displaymath}
 =\left\{ \begin{array}{ll}
 \pm 2 \eta' |a|\; |b| |\lan b'_N | b_N\ran| cos (\phi_a-\phi_b-\alpha_G) & \textrm{ if $k$ is even} \\

\pm 2 \eta |a|\; |b|  |\lan b'_N | b_N\ran|  sin (\phi_a-\phi_b-\alpha_G) & \textrm{ if $k$ is odd} \\
\end{array} \right.
\end{displaymath}

 $ t_Q=\pm \chi' a c^* \lan b_N|b''_N\ran \pm \chi' a^* c \lan b''_N | b_N\ran= \pm 2 \chi' |a| |c| |\lan b''_N | b_N\ran| cos (\phi_a-\phi_c-\alpha_Q)$, \\ 
  
 $t_{Q_k}=\pm (i)^k \chi' a c^*  \lan b_N|b''_N\ran \pm (-i)^k \chi' a^* c \lan b''_N | b_N\ran$ \\ 
  
 \begin{displaymath}
 =\left\{ \begin{array}{ll}
 \pm 2 \chi' |a|\; |c| |\lan b''_N | b_N\ran| cos (\phi_a-\phi_c-\alpha_Q) & \textrm{ if $k$ is even} \\

\pm 2 \chi |a|\; |c|  |\lan b''_N | b_N\ran|  sin (\phi_a-\phi_c-\alpha_Q) & \textrm{ if $k$ is odd} \\
\end{array} \right.
\end{displaymath}

$ t_Z=\pm \delta' b c^* \lan b'_N|b''_N\ran \pm \delta' b^* c \lan b''_N | b'_N\ran= \pm 2 \delta' |b| |c| |\lan b''_N | b'_N\ran| cos (\phi_b-\phi_c-\alpha_Z)$,  \\  
  
$t_{Z_k}=\pm (i)^k \delta' b c^*  \lan b'_N|b''_N\ran \pm (-i)^k \delta' b^* c \lan b''_N | b'_N\ran $\\  
 \begin{displaymath}
 =\left\{ \begin{array}{ll}
 \pm 2 \delta' |b|\; |c| |\lan b''_N | b'_N\ran| cos (\phi_b-\phi_c-\alpha_Z) & \textrm{ if $k$ is even} \\

\pm 2 \delta' |b|\; |c|  |\lan b''_N | b'_N\ran|  sin (\phi_b-\phi_c-\alpha_Z) & \textrm{ if $k$ is odd} \\
\end{array} \right.
\end{displaymath}

  Finally we get \\
   $|| \mathcal{T}^{(N)}_{|\psi\ran}||^2 = (\fr{3^{N-1}}{2^{N-1}})^2\times \fr{1}{3^N}\Big{[}(-2)^{\ell}|a|^2 +(-2)^{\ell'} |b|^2+(-2)^{\ell''}|c|^2)^2   +\sum^{N-1-s}_{k=1}\Big{(}\fr{(-2)^s}{3^{N-1-k}}\Big{)}^2 |a|^2 \binom{N-1-s}{k}$
$$+\sum^{N-1-s'}_{k=1}\Big{(}\fr{(-2)^{s'}}{3^{N-1-k}}\Big{)}^2|b|^2\binom{N-1-s'}{k}+\sum^{N-1-s''}_{k=1}\Big{(}\fr{(-2)^{s''}}{3^{N-1-k}}\Big{)}^2|c|^2\binom{N-1-s''}{k}$$ 
$$+ 4 \eta'^2 |a|^2 |b|^2 |\lan b'_N | b_N\ran|^2 cos^2 (\phi_a-\phi_b-\alpha_G) \sum_{k=0}^{m} \binom{m}{2k}$$
$$+4 \eta'^2 |a|^2 |b|^2 |\lan b'_N | b_N\ran|^2 sin^2 (\phi_a-\phi_b-\alpha_G) \sum_{k=0}^{m} \binom{m}{2k+1}$$
$$+ 4 \chi'^2 |a|^2 |c|^2 |\lan b''_N | b_N\ran|^2 cos^2 (\phi_a-\phi_c-\alpha_Q) \sum_{k=0}^{m'} \binom{m'}{2k}$$
$$+4 \chi'^2 |a|^2 |c|^2 |\lan b''_N | b_N\ran|^2 sin^2 (\phi_a-\phi_c-\alpha_Q) \sum_{k=0}^{m'} \binom{m'}{2k+1}$$
$$+ 4 \delta'^2 |b|^2 |c|^2 |\lan b''_N | b'_N\ran|^2 cos^2 (\phi_b-\phi_c-\alpha_Z) \sum_{k=0}^{m''} \binom{m''}{2k}$$
$$+4 \delta'^2 |b|^2 |c|^2 |\lan b''_N | b'_N\ran|^2 sin^2 (\phi_b-\phi_c-\alpha_Z) \sum_{k=0}^{m''} \binom{m''}{2k+1}\Big{]}$$ 
  where $m \le q$ is the number of $\lambda_1$ operators in $G$, $m' \le q'$ is the number of $\lambda_4$ operators in $Q$ and $m'' \le q''$ is the number of $\lambda_6$ operators in $Z$. Since $|\lan b'_N | b_N\ran|^2 \le 1$, $|\lan b''_N | b_N\ran|^2 \le 1$ and $|\lan b'_N | b''_N\ran|^2 \le 1$ we see that $$|| \mathcal{T}^{(N-1)}_{\rho}||^2 \le || \mathcal{T}^{(N)}_{|\psi\ran}||^2$$  
  
  equality occurring when $|b_N\ran =|b'_N\ran$ in which case $|\psi\ran=|\phi\ran\otimes|b_N\ran.$  
  It is straightforward, but tedious to elevate is proof for the general case 
  $$|\psi\ran=\sum_{\alpha_1 \cdots \alpha_N} a_{\alpha_1 \cdots \alpha_N} |b_{\alpha_1}\cdots b_{\alpha_N}\ran, \; \alpha_i=1,2,\cdots,8$$  
  Basically we have to keep track of $\binom{r}{2}$ $Y,\;B,\; E,\;D$ type of operators, where $r$ is the number of terms in the expansion of $|\psi\ran$, in order to obtain all nonzero elements of $\mathcal{T}^{(N)}_{|\psi\ran}$. When $N$th particle is traced out, the corresponding elements of $\mathcal{T}^{(N-1)}_{\rho}$ get multiplied by the overlap amplitudes, which leads to the required result. The generalization to qudits requires action of $\lambda$ operators on the computational basis . Rest of the proof has a straight forward generalization.\hfill $\blacksquare$\\

 \textit{Continuity of $E_{\mathcal{T}}$}: We show that for $N$-qudit pure states $||(|\psi\ran\lan\psi|-|\phi\ran\lan\phi|)||\rightarrow 0 \Rightarrow \Big{|}E_{\mathcal{T}}(|\psi\ran)-E_{\mathcal{T}}(|\phi\ran)\Big{|}\rightarrow 0 $
 
 \textit{Proof.} $||(|\psi\ran\lan\psi|-|\phi\ran\lan\phi|)||\rightarrow 0 $
$\Rightarrow ||\mathcal{T}^{(N)}_{|\psi\ran}-\mathcal{T}^{(N)}_{|\phi\ran}||\rightarrow 0 $

But  $||\mathcal{T}^{(N)}_{|\psi\ran}-\mathcal{T}^{(N)}_{|\phi\ran}|| \ge \Big{|}||\mathcal{T}^{(N)}_{|\psi\ran}||-||\mathcal{T}^{(N)}_{|\phi\ran}||\Big{|}$

Therefore $||\mathcal{T}^{(N)}_{|\psi\ran}-\mathcal{T}^{(N)}_{|\phi\ran}||\rightarrow 0 \Rightarrow \big{|}||\mathcal{T}^{(N)}_{|\psi\ran}||-||\mathcal{T}^{(N)}_{|\phi\ran}||\big{|} \rightarrow 0$
  
   $\Rightarrow \Big{|}E_{\mathcal{T}}(|\psi\ran)-E_{\mathcal{T}}(|\phi\ran)\Big{|}\rightarrow 0. $ \hfill $\blacksquare$\\

\subsection{ Entanglement of multiple copies of a given state}

{\it LU invariance. } We show that  $E_{\mathcal{T}}$ for multiple copies of $N$-qudit pure state $|\psi\ran$ is $LU$ invariant.
Consider a system of $N\times k$ qudits in the state $|\chi\ran=|\psi\ran\otimes |\psi\ran \otimes \cdots\otimes |\psi\ran$ ($k$ copies). It is straightforward to check that [34]

$$  \mathcal{T}^{(N)}_{|\chi\ran}= \mathcal{T}^{(N)}_{|\psi\ran}\circ \mathcal{T}^{(N)}_{|\psi\ran}\circ \cdots \circ \mathcal{T}^{(N)}_{|\psi\ran} \eqno{(22)}$$
   
 This implies, in a straightforward way, that $$|| \mathcal{T}^{(N)}_{|\chi\ran}||=||  \mathcal{T}^{(N)}_{|\psi\ran}||^k.$$
    Since by proposition 6 $||  \mathcal{T}^{(N)}_{|\psi\ran}||$ is $LU$ invariant, so is $|| \mathcal{T}^{(N)}_{|\chi\ran}||$.

 \textit{Superadditivity }: We have to show, for $N$qudit states   $|\psi\ran$ and $|\phi\ran$ that $$E_{\mathcal{T}}(|\psi\ran \otimes |\phi\ran )\ge E_{\mathcal{T}}(|\psi\ran)+ E_{\mathcal{T}}(|\phi\ran). \eqno{(23)}$$  
We already know that for $|\chi\ran=|\psi\ran\otimes|\phi\ran$
$$|| \mathcal{T}^{(N)}_{|\chi\ran}||= ||  \mathcal{T}^{(N)}_{|\psi\ran}||\; ||  \mathcal{T}^{(N)}_{|\phi\ran}||$$

Thus Eq. (23) gets transformed to $$||  \mathcal{T}^{(N)}_{|\psi\ran}||\; ||  \mathcal{T}^{(N)}_{|\phi\ran}||-(\fr{d(d-1)}{2})^N \ge ||  \mathcal{T}^{(N)}_{|\psi\ran}||+ ||  \mathcal{T}^{(N)}_{|\phi\ran}||-2(\fr{d(d-1)}{2})^{N/2}$$
 Putting $||  \mathcal{T}^{(N)}_{|\psi\ran}||=[(\fr{d(d-1)}{2})^{N/2}+x];\; x \ge 0$
 this inquality reduces to $(\fr{d(d-1)}{2})^{N/2} \ge 1-\fr{x}{2}$
which is satisfied by all $d \ge 2$ for all $x \ge 0$.

\subsection{Computational considerations}

Computation or experimental determination of $E_{\mathcal{T}}$ involves $(d^2-1)^{N}$ elements of $\mathcal{T}^{(N)}$ so that it increases exponentially with the number of qudits $N$. However, for many important classes of states, $E_{\mathcal{T}}$ can be easily computed and increases only polynomially with $N$. We have already computed $E_{\mathcal{T}}$ for the class of $N$ qutrit $GHZ$ states. For symmetric or antisymmetric states $\mathcal{T}^{(N)}$ is supersymmetric, that is, the value of its elements are invariant under any permutation of its indices [34]. This reduces the problem to the computation of $\fr{1}{7!}\Pi^7_{k=1}(N+k)$ distinct elements of $\mathcal{T}^{(N)}$ for qutrits, which is a polynomial in $N$[52].\\

\section{ Extension to mixed states}

The extension of $E_{\mathcal{T}}$ to mixed states $\rho$ can be made via the use of the {\it convex roof} or {\it (hull)} construction as was done for the entanglement of formation [16]. We define $E_{\mathcal{T}}(\rho)$ as a minimum over all decompositions $\rho=\sum_{i}p_i\arrowvert\psi_i\ran\lan\psi_i\arrowvert$ into pure states i.e.
$$E_{\mathcal{T}}(\rho)=\min_{\begin{subarray}{I} 
  \hskip  .1cm  {\{p_i,\psi_i\}}
 \end{subarray}}
\sum_{i}p_iE_{\mathcal{T}}(\arrowvert\psi_i\ran).\eqno{(24)}$$

The existence and uniqueness of the convex roof for $E_{\mathcal{T}}$ is guaranteed because it is a continuous function on the set of pure states [53]. This entanglement measure is expected to satisfy conditions (a), (b) and (c) given in section 4 and is expected to be  
(d)convex under discarding of information, i.e. $$\sum_{i}p_i E_{\mathcal{T}}(\rho_i)\geq E_{\mathcal{T}}(\sum_{i}p_i\rho_i). \eqno{(25)}$$ 

The criteria (a)-(d) above are considered to be the minimal set of requirements for any entanglement measure so that it is an entanglement monotone [44].

Evidently, criteria (a) and (b) are satisfied by $E_{\mathcal{T}}(\rho)$ defined via convex roof as it is satisfied by
$E_{\mathcal{T}}$ for pure states. Condition (d) follows from the fact that every convex hull (roof) is a convex function [54]. We need to prove (c), which is summarized in the following proposition.
 
\textit{Proposition 8}: If  a $N$-qudit mixed state $\rho$ is subjected to a local operation on $i$ th qudit giving outcomes $k$ with probabilities $p_k$ and leaving residual $N$ qudit mixed state $\rho_k$, then the expected entanglement $\sum_{k} p_k E_{\mathcal{T}}(\rho_k)$ of the residual state is not greater than the entanglement $E_{\mathcal{T}}(\rho)$ of the original state. $$\sum_{k} p_k E_{\mathcal{T}}(\rho_k) \le E_{\mathcal{T}}(\rho)$$ (If the operation is simply throwing away part of the system, then there will be only one value of $k$, with unit probability.)

The proof follows from the monotonicity of $E_{\mathcal{T}}(|\psi\ran)$  for pure states that is propositions 5,6 and 7.
Bennett et al. prove a version of proposition 8 in [48], which applies to any measure satisfying propositions 5,6 and 7. Thus the same proof applies to proposition 8, so we skip it. 

Note that any sequence of local operations comprises local operations drawn from the set of basic local operations (i)-(iv) stated in section 4, so that proposition 8 applies to any such sequence. Thus we can say that expected entanglement of a $N$-qudit system, measured by $E_{\mathcal{T}}(\rho)$, does not increase under local operations. \\

\section{Summary}

In summary we state that, we propose an entanglement measure (Eq. (8)) for $N$-qudit pure states which passes all the tests expected  of a good entanglement measure. Moreover, this measure is experimentally obtainable, without a prior knowledge of the state of the system, by measuring $\lambda$ operators ( generators of $SU(d)$) which are simply related to the angular momentum operators [54]. This measure can be computed efficiently, at lest for the state belonging to the symmetric or antisymmetric subspaces. \\

\textbf{Acknowledgments}

We thank Guruprasad Kar and Professor R. Simon for encouragement. We thank Sougato Bose for his helpful suggestion. A.S.M.H. thanks Sana'a University for financial support

\textbf{References}\\

\begin{verse} 

[1] C. H. Bennett, G. Brassard, C. Cr\'epeau, R. Jozsa, A. Peres,
and W. K. Wootters, Phys. Rev. Lett. \textbf{70}, 1895 (1993).

[2] Ye Yeo and Wee Kang Chua, Phys. Rev. Lett. \textbf{96}, 060502 (2006).

[3] C. H. Bennett and G. Brassard (unpublished); D. Deutsch, A. Ekert, R. Jozsa, C. Macchiavello, S. Popescu, and A. Sanpera, Phys. Rev. Lett. \textbf{77} 2818, (1996), \textbf{80}, 2022 (1998), H.-K. Lo, {\it in Introduction to Quantum Computation and Information}, edited by H.-K. Lo, S. Popescu and T. Spiller (World Scientific,
Singapore, 1998), pp. 76–119; H. Zbinden {\it ibid.} pp. 120–142.

[4] C. H. Bennett and S. J. Wiesner, Phys. Rev. Lett. 69, 2881 (1992).

[5] C. H. Bennett, C. A. Fuchs, and J. A. Smolin, in {\it Quantum
Communication, Computing and Measurement}, edited by O. Hirota, A. S. Holevo, and C. M. Caves (Plenum, New York, 1997).

[6] C. H. Bennett, P. W. Shor, J. A. Smolin, and A. V. Thapliyal,
Phys. Rev. Lett. \textbf{83}, 3081 (1999).

[7] P. W. Shor, Phys. Rev. A \textbf{52}, R2493 (1995); D. Gottesman,
Ph.D. Thesis, California Institute of Technology, 1997; LANL
e-print, quant-ph/9705052.

[8] D. Deutsch, Proc. R. Soc. London, Ser. A \textbf{400}, 97 (1985); D.
Deutsch ibid., \textbf{425}, 73 (1989); M. A. Nielsen and I. L. Chuang, {\it Quantum Computation and Quantum Information}, (Cambridge University Press 2000).

[9] L. K. Grover, LANL e-print, quant-ph/9704012.

[10] R. Cleve and H. Buhrman, LANL e-print quant-ph/9704026.

[11] T. J. Osborne and M. A. Nielsen, Phys. Rev. A  \textbf{66}, 032110 (2002); A. Osterloh et al., Nature (London)  \textbf{416}, 608 (2002).

[12] M. B. Plenio, S. Virmani, Phys. Rev. Lett. \textbf{99}, 120504 (2007)

[13] M. B. Plenio and S. Virmani , Quantum Inf. Comput., Vol.  \textbf{7}, 1 (2007).

[14] K. \.Zyczkowski and I. Bengstsson, quant-ph/0606228.

[15] R., P., M., K. Horodecki, quant-ph/0702225v2.

[16] W.K. Wootters, Phys. Rev. Lett. \textbf{80}, 2245 (1998).

[17] K. G. H. Vollbrecht and R.F. Werner, Phys. Rev. A \textbf{64}, 062307 (2001).

[18] B.M. Terhal and K.G.H. Vollbrecht, Phys. Rev. Lett. \textbf{85}, 2625 (2000).

[19] A. R. Usha Devi, R. Prabhu, and A. K. Rajagopal, Phys. Rev. Lett.   \textbf{98}, 060501 (2007).\\

[20] D. Kaslikowski et al., Phys. Rev. Lett. \textbf{85}, 4418 (2000).\\

[21] D. Collins et al., Phys. Rev. Lett. \textbf{88}, 040404 (2002).\\

[22] H. Bechmann-Pasquinucci and A. Peres, Phys. Rev. Lett. \textbf{85}, 3313 (2000).\\

[23] T. Durt, N. J. Cerf, N. Gisin, and M. Zukowski, Phys. Rev. A \textbf{67}, 012311 (2003).\\

[24] M. Bourennane, A. Karlsson, and G. Bjork, Phys. Rev. A \textbf{64}, 012306 (2001).\\

[25] N. J. Cerf, M. Bourennane, A. Karlsson, and N. Gisin, Phys. Rev. Lett. \textbf{88}, 127902
(2002).\\

[26] J. C. Howell, A. Lamas-Linares, and D. Bouwmeester, Phys. Rev. Lett. \textbf{85}, 030401
(2002).\\

[27] R. T. Thew, A. Ac´in, H. Zbinden and N. Gisin, Phys. Rev. Lett. \textbf{93}, 010503 (2004).\\

[28] A. Vaziri, G. Weihs, and A. Zeilinger, Phys. Rev. Lett. \textbf{89}, 240401 (2002).\\

[29] H. de Riedmatten, I. Marcikic, H. Zbinden and N. Gisin, Quant. Inf. and Comp. \textbf{2},
425 (2002).

[30] V. Vedral and M.B. Plenio, Phys. Rev. A \textbf{57}, 1619 (1998).

[31] K. \.Zyczkowski, P. Horodecki, A. Sanpera, and M. Lewenstein,
Phys. Rev. A \textbf{58}, 883 (1998); G. Vidal and R.F. Werner, ibid.
\textbf{65}, 032314 (2002).

[32]  J. Eisert and H.J. Briegel, Phys. Rev. A \textbf{64}, 022306 (2001).

[33] David A. Meyer and Nolan R. Wallach, J. Math. Phys.  \textbf{43}, 4273 (2002).

[34] Ali Saif M. Hassan and Pramod S. Joag, Quantum Inf. Comput., Vol.  \textbf{8}, 0773 (2007).

[35] F. Bloch , Phys. Rev. \textbf{70}, 460 (1946).

[36] G. Kimura and A. Kossakowski, Open Sys. Inf. Dyn. \textbf{12}, 207 (2005).

[37] G. Kimura,  Phys. Lett. A \textbf{314}, 339 (2003).

[38] M.S. Byrd and N. Khaneja,  Phys. Rev. A \textbf{68}, 062322 (2003).

[39] G. Mahler and V.A. Weberru\ss, {\it Quantum Networks}, (Springer , Berlin, 1995).

[40]  Julio I. de Vicente, Quantum Inf. Comput. \textbf{7}, 624 (2007).

[41] J. E. Harriman, Phys. Rev. A \textbf{17}, 1249 (1978).

[42] A. Kossakowski, Open Sys. Inf. Dyn. \textbf{10}, 213 (2003).

[43] Kai Chen, Sergio Albeverio and Shao-Ming Fei, Phys. Rev. Lett. \textbf{95}, 040504 (2005).

[44] G. Vidal, J. Mod. Opt.  \textbf{47}, 355 (2000).

[45] I. Bengtsson and K. \.Zyczkowski, {\it Geometry of Quantum States}, (Cambridge University Press 2006).

[46] T. G. Kolda, {\it Multilinear operators for higher-order decompositions}, Tech. Report SAND2006-2081, Sandia National Laboratories, Albuquerque, New Mexico and
Livermore, California, Apr. 2006.

[47] T. G. Kolda,  SIAM J. Matrix Anal. A., Vol. \textbf{23}, 243 (2001).

[48] C. Bennett, D. P. DiVincenzo, J. A. Smolin, and W. K. Wootters, Phys. Rev.  A \textbf{54}, 3824 (1996).

[49] L. De Lathauwer, B. De Moor, and J. Vandewalle, SIAM J. Matrix Anal. A., Vol. \textbf{21}, 1253 (2000).
 
[50] Jing Zhang, Chun-Wen Li, Jian-Wu Wu, Re-Bing Wu and Tzyh-Jong Tam, Phys. Rev. A, \textbf{73}, 022319 (2006).
 
[51] R.A. Horn and C.R. Johnson, {\it  Matrix Analysis}, Cambridge University Press (Cambridge 1985).

[52] William Feller, {\it An Introduction to Probability Theory and it Applications}, vol. I, (Wiley Eastern University Edition, 1983)

[53] A. Uhlmann, Phys. Rev. A \textbf{62}, 032307 (2000).

[53] A. Uhlmann, LANL e-print, quant-ph/9704017v2.
  
[54] R. A. Bertlmann and P. Krammer,  arXiv: 0806.1174v1 [quant-ph]; I. P. Mendas, J. Phys. A \textbf{39}, 11313 (2006).
  
\end{verse}

\end{document}